\documentclass[aps,twocolumn,showpacs]{revtex4}
\usepackage[dvips]{graphicx}

\begin{document}

\title{Flux induced semiconducting behavior of a quantum network} 

\author{Shreekantha Sil$^1$, Santanu K. Maiti$^2$ and Arunava Chakrabarti$^3$}

\affiliation{$^1$Department of Physics, Visva-Bharati, Santiniketan, 
West Bengal-731 235, India. \\
$^2$Department of Physics, Narasinha Dutt College, 129 Belilious Road, 
Howrah-711 101, India. \\
$^3$Department of Physics, University of Kalyani, Kalyani, 
West Bengal-741 235, India.} 

\begin{abstract}
We show that a diamond shaped periodic network, recently proposed 
as a model of a spin filter (A. Aharony {\em et al.}~\cite{aharon08}) 
is capable of behaving as a p-type or an n-type semiconductor depending 
on a suitable choice of the on-site potentials of the atoms occupying 
the vertices of the lattice, and the strength of the magnetic flux 
threading each plaquette of the network. A detailed study of the density 
of states of an infinite network is made together with the conductance 
of finite sized system to establish the idea.
\end{abstract}


\maketitle 

\noindent
Low dimensional model quantum systems have been the objects of intense 
research, both in theory and in experiments, mainly due to the fact that 
these simple looking systems are prospective candidates for nano devices 
in electronic as well as spintronic engineering~\cite{ aharon08,ahar2,
ore1,ore2,levy,torio,buks96,fuhrer07,kob,popp03,foldi08,berc04,berc2}. 
Apart from this feature, several striking spectral properties are 
exhibited by such systems owing to the quantum interference which 
is specially observed in quantum networks containing closed loops. 
Examples are, the Aharonov-Bohm (AB) effect~\cite{levy} in the 
magnetoconductance of quantum dots~\cite{buks96}, electron 
transport in quantum dot arrays~\cite{ore1,ore2}, Fano effect in 
a quantum ring-quantum dot system~\cite{fuhrer07}, spin filter 
effects in mesoscopic rings~\cite{popp03,foldi08} and 
dots~\cite{torio}, to name a few.

Recently, Aharony {\em et al.}~\cite{aharon08,ahar2} have proposed a model 
of a nano spintronic device using a linear chain of diamond-like blocks of 
atomic sites. Each plaquette of the array is threaded by identical magnetic 
flux. They have analyzed how the Rashba spin orbit interaction and the AB 
flux combine to select a propagating ballistic mode. A similar chain was 
earlier investigated by Bercioux {\em et al.}~\cite{berc04,berc2} in 
the context of spin polarized transport of electrons. However, there are 
certain special spectral features offered by the diamond chain, 
particularly the role of the AB flux, which we believe, remain 
unexplored. This is precisely the area we wish to highlight in the 
present communication. We show that an infinite diamond chain of 
identical atoms behaves as an insulator at $T=0$ K in the presence 
of a non-zero AB flux. As we arrange atoms of two different kinds 
(represented by two different values of the on-site potential) 
periodically on a diamond chain, a highly degenerate localized 
level is created near one of the two sub-bands of extended states. The 
proximity of this localized level to either of the sub-bands can be 
controlled by tuning the AB flux, and can be made to stay arbitrarily 
close to either of the sub-bands. The entire system is then capable 
of behaving as an $n$-type or a $p$-type semiconductor as explained 
later. The conductance spectrum of a finite array of the diamond 
plaquettes is also studied to judge the applicability of such a 
network geometry in device engineering.

\begin{figure}[ht]
{\centering\resizebox*{7.5cm}{5cm}{\includegraphics{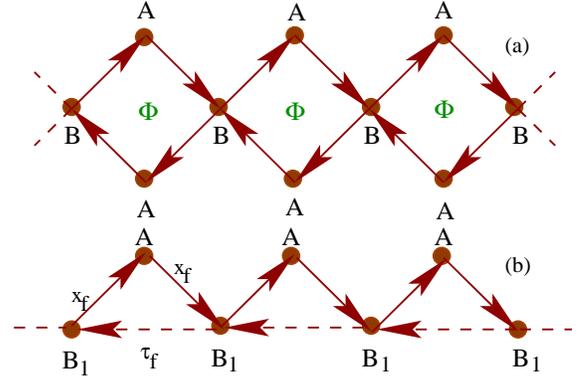}} \par}
\caption{(a) Schematic view of a section of an infinite diamond chain. 
(b) The RG scheme. The arrow gives 
the direction of the {\em forward} hopping $t\exp(i\theta_{ij})$.}
\label{bridge1}
\end{figure}

We adopt a tight binding formalism, and incorporate only the nearest 
neighbor hopping. We begin by referring to Fig.~\ref{bridge1}(a). 
A magnetic (AB) flux $\phi$ threads each plaquette. The Hamiltonian 
of the network is given by, 
\begin{equation}
H=\sum_i \epsilon_i c_i^{\dagger} c_i + \sum_{<ij>} t
\left(c_i^{\dagger} c_j e^{i\theta_{ij}} + c_j^{\dagger} c_i 
e^{-i\theta_{ij}}\right)
\label{equ1}
\end{equation}
where, 
$c_{i}$ ($c_{i}^{\dagger}$) are the annihilation (creation) operator at the 
$i$th site of the network, $\epsilon_i$ is the on-site potential at the 
$i$-th site which we shall choose as $\epsilon_A$ or $\epsilon_B$ as shown, 
and $t$ is the constant nearest neighbor hopping integral. The phase 
$\theta_{ij}$ is constant, and is given by, 
$\theta_{ij} = \pm 2\pi\phi/4\phi_0$, where $\phi_0=hc/e$ is the flux 
quantum. The positive or negative sign of the phase factor depends 
on whether the electron hops from site $A(B)$ to site $B(A)$ {\it along 
the arrow} (called the `forward' hopping) or against it (`backward' 
hopping).

To obtain the average density of states (AVDOS) of the infinite system we 
make use of the system of equations satisfied by the Green's function 
$G_{ij}$, viz, 
\begin{equation}
(E-\epsilon_i) G_{ij} = \delta_{ij} + t \sum_k e^{i\theta_{ik}} G_{kj}
\label{equ5}
\end{equation}
On the right hand side of Eq.~\ref{equ5} the index $k$ runs over the 
nearest neighbors of the $i$-th site. The density of states is obtained 
by a decimation renormalization group (RG) method~\cite{south83}.
The RG scheme is depicted in Fig.~\ref{bridge1}. The lower $A$-vertices in 
Fig.~\ref{bridge1}(a) are decimated first, and the original diamond 
array gets transformed into an array of triangular plaquettes with 
the surviving $A$ atoms sitting at the top  and the pair of original 
$B$-atoms getting renormalized into a pair of $B_1$ atoms, each 
characterized by the effective on-site potential 
$\epsilon_{B_1}=\epsilon_B+2t^2/(E-\epsilon_A)$. The $B_1$-$B_1$ effective 
hopping integral is $\tau_f=t^2 \exp(i\pi\phi/\phi_0)/(E-\epsilon_A)$
and its complex conjugate $\tau_b=\tau_F^{*}$. 
$x_f=t \exp(i\pi \phi/2\phi_0)$ 
is the $B_1$-$A$ hopping integral as shown in Fig.~\ref{bridge1}(b). 
Its conjugate is named $x_b$. The subscripts $'f'$ and $'b'$ refer 
to the {\it forward} or {\it backward} hopping due to the broken 
time reversal symmetry due to the magnetic flux. In Fig.~\ref{bridge1}(b) 
we have not shown the flux explicitly as the flux has been automatically 
included in $x_f$, $x_b$, $\tau_f$ and $\tau_b$. The array of triangles 
is now renormalized by decimating the alternate $A$ sites and the 
appropriate $B$-sites preserving the triangular geometry of the array. 
The recursion relations satisfied by the parameters are given by, 
\begin{eqnarray}
\epsilon_A' & = & \epsilon_A + p_2x_b + p_1x_f \nonumber \\
\epsilon_{B_1}' & = & \epsilon_{B_1} + 2\frac{x_fx_b}{E-\epsilon_A} + 
h_fr_1 + h_br_2 
\end{eqnarray}
for the site potentials and, 
\begin{eqnarray}
x_f'= h_b~p_2~;~~ \tau_f' = h_f~q_1
\label{equ4}
\end{eqnarray}
for the hopping integrals between $B_1$-$A$ and $B_1$-$B_1$ pairs 
on the renormalized triangular array. Obviously, $x_b^{\prime}$ and 
$\tau_b^{\prime}$ are given by, $x_f^{\prime,*}$ and $\tau_f^{\prime,*}$ 
respectively, and, $p_1=[x_b(E-\epsilon_2) + x_f\tau_f]/\delta$, 
$p_2=p_1^{*}$, $h_f=\tau_f + x_b^{2}/(E-\epsilon_A)$, $h_b=h_f^{*}$, 
$r_1=[h_b(E-\epsilon_2)]/\delta$, $r_2=r_1^{*}$ with 
$\epsilon_2=\epsilon_{B_1} + x_fx_b/(E-\epsilon_A)$ and 
$\delta=(E-\epsilon_2)^2 - \tau_f \tau_b$. With a small imaginary 
part added to the energy $E$, the local Green's functions at the 
sites $A$ and $B$ of the original diamond array are then obtained as, 
\begin{eqnarray}
G_{AA} = \frac{1}{E+i0^{+}-\epsilon_A^{*}}~;~~
G_{BB}= \frac{1}{E+i0^{+}-\epsilon_{B_1}^{*}}
\end{eqnarray}
where, the superscript $*$ above refer to the fixed point values of the 
respective parameters. The AVDOS is given by
\begin{equation}
\rho(E)=\frac{2}{3}\rho_A(E)+\frac{1}{3}\rho_B(E)
\end{equation}
where $\rho_i(E)=-(1/\pi) Im~ G_{ii}$, $i$ being $A$ or $B$.

\begin{figure}[ht]
\begin{center}
\includegraphics[height=15cm,width=7cm,angle=270]{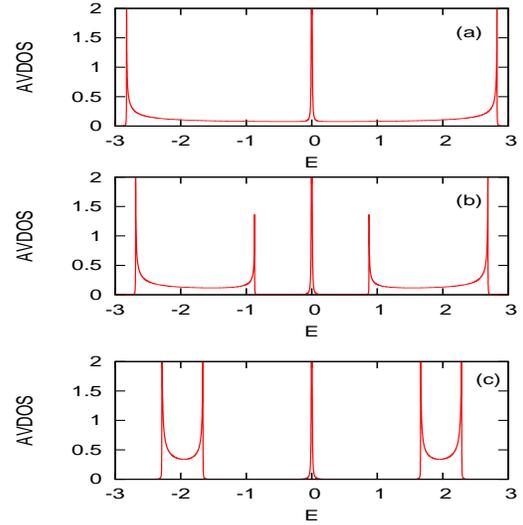}
\caption{$\rho(E)$-$E$ for an infinite diamond network array when the AB 
flux is (a) $\phi=0$, (b) $\phi=\phi_0/5$, and (c) $\phi_=2\phi_0/5$. We 
have chosen $\epsilon_A = \epsilon_B = 0$ and $t=1$. The AVDOS has been 
plotted within the range zero to two.}
\label{dos1}
\end{center}
\end{figure}

\begin{figure}[ht]
\begin{center}
\includegraphics[height=15cm,width=7cm,angle=270]{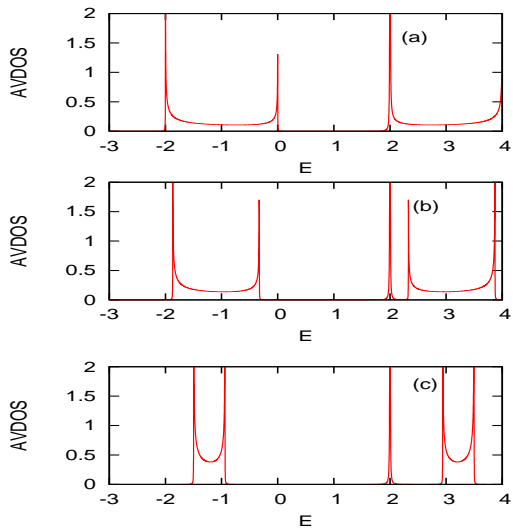}
\caption{$\rho(E)$-$E$ for an infinite diamond network array when the AB 
flux is (a) $\phi=0$, (b) $\phi=\phi_0/5$, and (c) $\phi_=2\phi_0/5$. We 
have chosen $\epsilon_A=2$ and $\epsilon_B=0$ with $t=1$. The AVDOS has 
been plotted within the range zero to two.}
\label{dos2}
\end{center}
\end{figure}

Let us now present two separate cases which will throw light on the 
central problem addressed in this paper, viz, the semiconducting 
behavior of such a network.
\vskip 0.1cm
\noindent
{\em Case I:} $\epsilon_A = \epsilon_B$ 
\vskip 0.1cm
\noindent
The AVDOS in this case is illustrated for $\epsilon_A=\epsilon_B=0$ 
in Figs.~\ref{dos1}(a), (b) and (c) for $\phi=0$, $\phi=\phi_0/5$ and 
for $\phi=2\phi_0/5$ respectively. In the zero flux case the spectrum 
is a continuum with a very high value of the density of states at the 
center, i.e. at $E=0$. The system exhibits a metallic character. As 
the magnetic field through the plaquette assumes non-zero value, a 
gap opens at the centre with the peak in the AVDOS fixed at $E=0$. 
This high value of the AVDOS corresponds to strongly localized states. 
It is to be appreciated that these states are localized strictly on 
the $A$-type vertices of the array and the care has to be taken while 
renormalizing the diamond array so that both the $A$-type sites do not 
get decimated. This precise RG scheme is explained earlier. If the 
state corresponding to an energy $E$ is localized, then the renormalized 
hopping integral will iterate to zero. This is observed for the central 
peak in the AVDOS by iterating Eq.~\ref{equ4}.
The gap around the central peak widens as the flux is increased gradually, 
and the sub-bands at the two extremities shrink to two sharp lines of zero 
width as $\phi=\phi_0/2$. The picture reverses as $\phi$ increases from 
$\phi_0/2$ to $\phi_0$. The exactly symmetrical location of the localized 
level with respect to the sub-bands at the flanks makes the system behave as 
a semiconductor when the electronic filling of the system lies between $1/3$ 
and $2/3$. More precisely, one can get a $p$-type semiconductor when the 
filling factor $n_e$ is $1/3$ and, an $n$-type semiconductor when 
$1/3 < n_e \le 2/3$. 
\vskip 0.1cm 
\noindent
{\em Case II:} $\epsilon_A \ne \epsilon_B$
\vskip 0.1cm
\noindent
The AVDOS is exhibited in Fig.~\ref{dos2} with $\epsilon_B=0$ and 
$\epsilon_A=2$ for different values of the flux. By comparing 
Fig.~\ref{dos1} and Fig.~\ref{dos2} we see that the highly 
degenerate localized level is pinned at $E=\epsilon_A$. This 
has been verified using various values of $\epsilon_A$ keeping 
$\epsilon_B$ fixed. So, the states are localized at the vertices 
$A$ of the diamond chain. The other interesting feature is that 
as the gaps open up for non-zero flux, the localized level 
is placed asymmetrically with respect to the continuous sub-bands of 
extended eigenstates. With a proper tuning of the flux, one such 
sub-band can be brought arbitrarily close to the sharp localized 
level. For example, with $\epsilon_A=2$, the localized level resides 
closer to the upper sub-band (the conduction band) than the lower 
one (the valence band). A reversal 
of the sign of the on-site potential $\epsilon_A$ reverses the picture. In 
either case, the localized level can be placed arbitrarily close to any one 
\begin{figure}[ht]
{\centering \resizebox*{7.25cm}{3.5cm}{\includegraphics{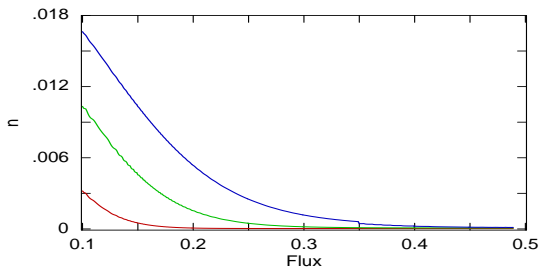}}\par}
\caption{Variation of the electron concentration $n$ as a function of
the flux $\phi$. The red, green and blue curves correspond to the
temperatures $k_{B}T=0.5$ , $1$  and $1.5$  respectively, $k_B$
being the Boltzmann constant.}
\label{electron}
\end{figure}
of the sub-bands by appropriately tuning the value of the magnetic flux. 
This control over the proximity of the degenerate localized level and 
a sub-band can be utilized to simulate an extrinsic semiconductor like 
behavior. For example, let us refer to the case in which $\epsilon_A=2$ 
and $\epsilon_B=0$, 
and fix the Fermi level at $E=2$ (i.e. all states from the bottom of the 
left (valence) band up to $E=2$ are filled up at $T=0$ K). The strongly 
degenerate level is pinned at $E=2$. If the magnetic field is small, 
then the energy gap between the localized level at $E=2$ and the bottom 
of the conduction band (the right sub-band) is small 
\begin{figure}[ht]
{\centering\resizebox*{8.5cm}{2.5cm}{\includegraphics{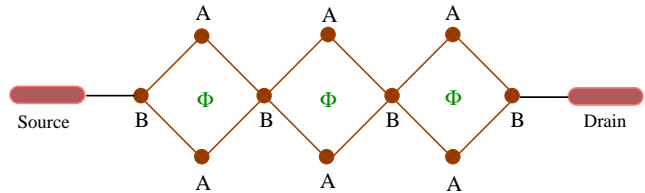}} \par}
\caption{Schematic view of a finite diamond chain attached
to two metallic electrodes.}
\label{diamond}
\end{figure}
enough for electrons to bridge. The system now behaves as an $n$-type 
semiconductor. By reverting to the case of $\epsilon_A=-2$ we can simulate 
a $p$-type semiconductor-like behavior observed by the same diamond chain. 
In this case the localized level is pinned at $E=-2$ and, we need to 
fix the Fermi level at the top of the valence band, so that electrons 
can jump into unoccupied levels above creating holes in the valence band. 
\begin{figure}[ht]
{\centering
\resizebox*{7.5cm}{8cm}{\includegraphics{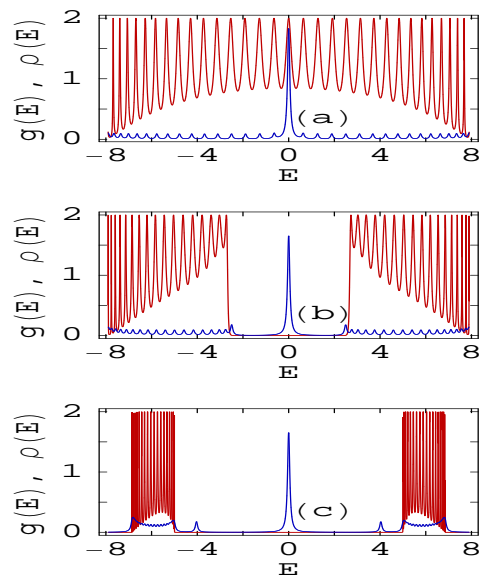}} \par}
\caption{$g$-$E$ (red color) and $\rho$-$E$ (blue color) curves for a
$20$-ring chain. (a) $\phi=0$, (b) $\phi=0.2$, and (c) $\phi=0.4$. Other
parameters are, $\epsilon_A=\epsilon_B=0$, $t=3$, $\epsilon_0=0$, $t_0=4$ 
and $\tau_S=\tau_D=2.5$.}
\label{cond1}
\end{figure}
It should be noted however, that with increasing flux the gap-width 
increases and hence the probability of electrons crossing over to any 
one of the sub-bands decreases. 

In Fig.~\ref{electron} we present the variation of the concentration of 
electrons (holes) at finite temperatures in the conduction (valence) 
bands for $\epsilon_A=2$ and $-2$ respectively. With flux increasing 
from zero the carrier concentration (electron or hole) diminishes.

The applicability of the above physics depends on whether such features 
are exhibited by systems with a finite size as well. In order to check 
this we have calculated the density of states and conductance of a twenty 
($20$) plaquette chain. To calculate the conductivity, the system is 
\begin{figure}[ht]
{\centering
\resizebox*{7.5cm}{8cm}{\includegraphics{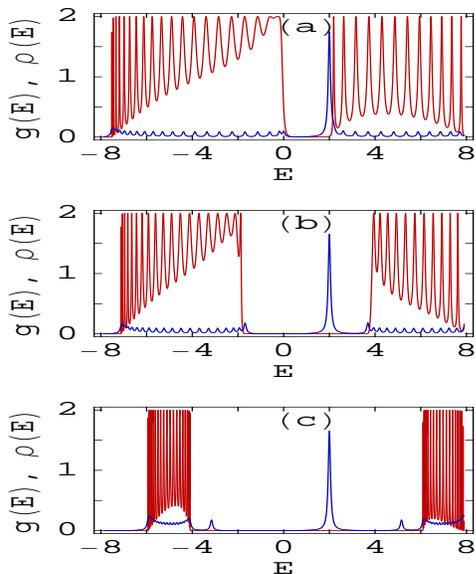}} \par}
\caption{$g$-$E$ (red color) and $\rho$-$E$ (blue color) curves for a
$20$-ring chain. (a) $\phi=0$, (b) $\phi=0.2$, and (c) $\phi=0.4$. Other
parameters are, $\epsilon_A=2$, $\epsilon_B=0$, $t=3$, $\epsilon_0=0$, 
$t_0=4$ and $\tau_S=\tau_D=2.5$.}
\label{cond2}
\end{figure}
connected between two metallic electrodes, viz, the source and the drain
(Fig.~\ref{diamond}), described by the standard tight binding Hamiltonian 
and parametrized by a constant site potential $\epsilon_0$ and nearest 
neighbor hopping integral 
$t_0$. The hopping integral between the source and the system is $\tau_S$,
while it is $\tau_D$ between the system and the drain.
Throughout the calculation we choose the units where $c=e=h=1$.
For low bias and low temperature one calculates the conductivity 
using the single channel Landauer formula \cite{datta} $g = (2e^2/h) 
\mathcal{T}$, where, the transmission coefficient $\mathcal{T}$ is given 
by~\cite{datta} $\mathcal{T} = Tr [\Gamma_S G^r \Gamma_D G^a]$. 
$\Gamma_S$ and $\Gamma_D$ correspond to the imaginary parts of the 
self energy due to the coupling of the diamond chain with the electrodes, 
and $G$ represents the usual Green's function. In Figs.~\ref{cond1} and 
\ref{cond2} we have shown simultaneously the AVDOS and the conductance 
spectra. Fig.~\ref{cond1} illustrates the symmetric case with 
$\epsilon_A=\epsilon_B=0$. 
With zero flux, the system exhibits oscillating conductance profile with 
resonance peaks. As the flux becomes non-zero, and increases in value, the 
conductance windows occupy the regions corresponding to the two sub-bands 
at the flanks, and shrink in width as the flux increases towards the half 
flux quantum. Figure~\ref{cond2} depicts a similar qualitative feature, 
but now with an asymmetric conductance profile as the potentials 
$\epsilon_A$ and $\epsilon_B$ are different.

Before we end, it may be mentioned that whether an array of such 
plaquettes will behave as an insulator or will have a metallic 
character in the absence of an external magnetic field, depends on the 
geometry of the unit cells. For example, with an array of identical 
triangular plaquettes a gap already exists in the middle of the spectrum 
even when the external magnetic field is zero. Thus such a triangular 
array will have an insulating character at $T=0$ K and at zero flux. 
However, by changing the site potentials one can generate strictly 
localized levels inside the gap and can again have the $n$- or $p$-type 
semiconducting behavior as discussed earlier in respect of the 
diamond plaquette-array. 

Finally, in view of the potential application of the networks as a 
device we would like to point out that the qualitative features 
presented here should remain valid even at finite temperature 
($\sim 300$ K) since, the broadening of the energy levels of the 
diamond array due to its coupling with the electrodes will be much 
larger than that of the thermal broadening~\cite{datta}.

\end{document}